# Different roles of Fe$_{1-x}$Ni$_x$OOH co-catalyst on hematite (α-Fe$_2$O$_3$) photoanodes with different dopants

Anton Tsyganok, Dino Klotz, Kirtiman Deo Malviya, Avner Rothschild, and Daniel A Grave*

Department of Materials Science and Engineering, Technion - Israel Institute of Technology, Haifa, Israel
*Email: dgrave@technion.ac.il

**Abstract**

Transparent Fe$_{1-x}$Ni$_x$OOH overlayers (~2 nm thick) were deposited photoelectrochemically on (001) oriented heteroepitaxial Sn- and Zn-doped hematite (α-Fe$_2$O$_3$) thin film photoanodes. In both cases, the water photo-oxidation performance was improved by the co-catalyst overlayers. Intensity modulated photocurrent spectroscopy (IMPS) was applied to study the changes in the hole current and recombination current induced by the overlayers. For the Sn-doped hematite photoanode, the improvement in performance after deposition of the Fe$_{1-x}$Ni$_x$OOH overlayer was entirely due to reduction in the recombination current, leading to a cathodic shift in the onset potential. For the Zn-doped hematite photoanode, in addition to a reduction in recombination current, an increase in the hole current to the surface was also observed after the overlayer deposition, leading to a cathodic shift in the onset potential as well as an enhancement in the plateau photocurrent. These results demonstrate that Fe$_{1-x}$Ni$_x$OOH co-catalysts can play different roles depending on the underlying hematite photoanode. The effect of the co-catalyst is not always limited to changes in the surface properties, but also to an increase in hole current from the bulk to the surface that indicates a possible crosslink between surface and bulk processes.

**Manuscript**

Hematite (α-Fe$_2$O$_3$) is an attractive material for solar water splitting based on its favorable properties as a photoanode material in photoelectrochemical (PEC) cells.[1] However, the performance of state-of-the-art hematite photoanodes[2–5] is still far short of the maximum theoretical efficiency, both in terms of photocurrent and photovoltage. One route for improving photoanode performance is through use of various co-catalysts which reduce the overpotential for water photo-oxidation, thereby leading to a cathodic shift in the applied bias.[6–9] One of the most promising materials for use as a co-catalyst is earth abundant Fe$_{1-x}$Ni$_x$OOH. For the remainder of the manuscript, we will refer to Fe$_{1-x}$Ni$_x$OOH as "FeNiO$_x$", a commonly used abbreviation. FeNiOx overlayers have shown similar improvements in photoelectrochemical performance as more expensive IrO$_x$ based co-catalysts.[10] FeNiO$_x$ overlayers can be produced easily by a variety of methods[11–15] and they are stable in alkaline solutions,[12] as the oxyhydroxide phase Fe$_{1-x}$Ni$_x$OOH.[16] In addition, it has recently been shown[17] that using a photoelectrochemical deposition method, very thin and transparent FeNiO$_x$ overlayers can be deposited to avoid optical (absorption) losses in the photoanode. Significant cathodic shifting of the onset potential for water photo-oxidation is typically observed for FeNiO$_x$ coated hematite photoanodes.[18,19] Generally, the changes in performance have been attributed to a reduction in the surface recombination either as a result of surface passivation,[20] hole storage in the overlayer,[21] or p-n junction formation.[22] For ultrathin photoelectrodeposited FeNiO$_x$ overlayers, improved catalysis has been suggested as the reason for improvement.[17] In some rare cases, improvement in bulk charge separation efficiency that gives rise to increased hole current to the surface has also been reported.[23] Despite significant study of this material, the underlying mechanisms for improvement in photoelectrochemical performance are not clear,[24] and comparisons between different FeNiO$_x$ coated hematite photoanodes prove difficult due to the different techniques used to fabricate both the co-catalyst and hematite layers, especially in nanostructured electrodes where changes in the layer fabrication may have significant effect on the photoanode morphology.[5] In this work, using well-controlled model heteroepitaxial hematite films[25] examined by an intensity modulated photocurrent spectroscopy (IMPS) analysis,[26] we show that the same FeNiO$_x$ overlayer can have different roles in improving the photoelectrochemical performance of the photoanode depending on the doping of the underlying hematite photoanode. Two dopants that were investigated in this study are Sn and Zn which are considered as donor and acceptor dopants, respectively.[27] In the case of Sn-doped hematite photoanodes, the improvement was found to result from reduced recombination, whereas for Zn-doped hematite photoanodes both reduction in recombination was observed as well as increase in the hole current to the surface. These results show that the role of the co-catalyst overlayer is strongly dependent on the underlying photoanode.

As a first approach, a photoelectrochemical deposition method for FeNiO$_x$ reported by Morales-Guio et al[17] was modified to a constant current photoelectrodeposition. The deposition was easily adapted on different types of hematite photoanodes: polycrystalline films deposited by ultrasonic spray pyrolysis (USP) and pulsed laser deposition (PLD) on fluorine doped tin oxide (FTO) coated glass substrates, as well as ultraflat heteroepitaxial hematite films deposited by PLD on transparent Nb-doped SnO$_2$ (NTO) and reflective Pt coated sapphire substrates. Schematics of the samples are presented in Figure S1. Figure 1(a) shows a transmission electron microscopy (TEM) lattice image of a cross-sectional lamella obtained from one of the FeNiO$_x$ coated hematite films deposited by PLD, showing conformal coating with an overlayer thickness of ~2 nm. The hematite is highly crystalline with a thin layer of FeNiO$_x$ at the surface. The EDS elemental maps presented in Figure 1(b) confirm the presence of the FeNiO$_x$ overlayer. More details regarding the fabrication of the photoanodes as well as their crystallinity and morphology are discussed in the SI.

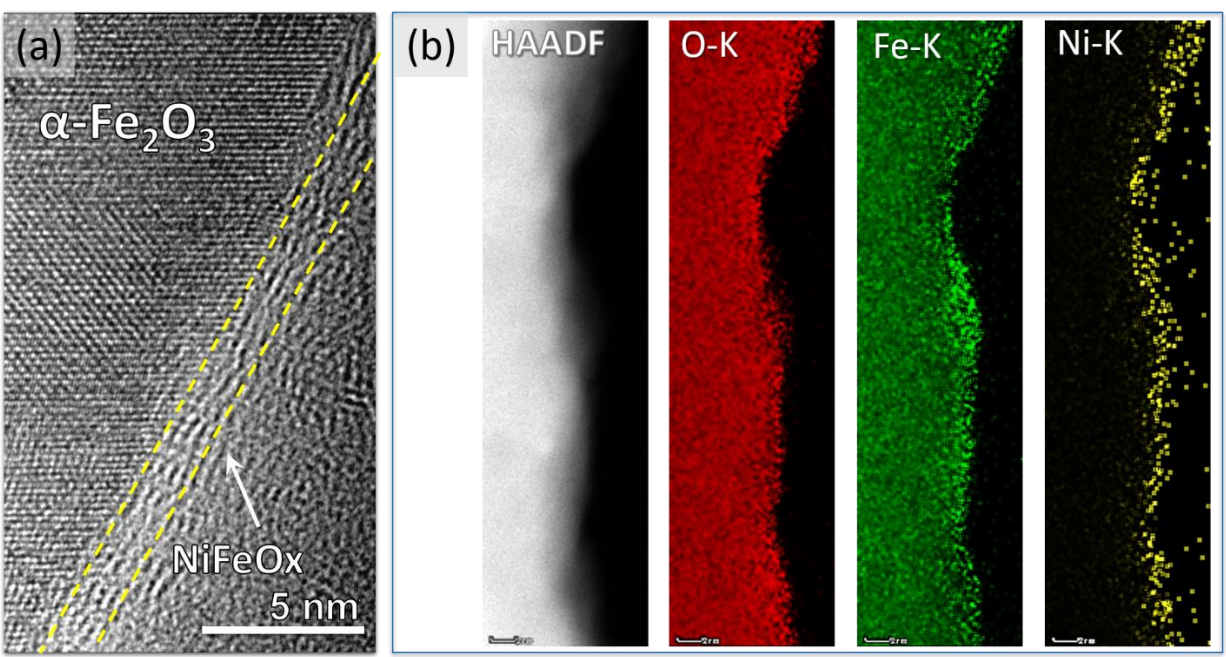

**Figure 1: (a)** High resolution TEM micrograph of the hematite photoanode surface. **(b)** STEM-HAADF image of the hematite with the EDS elemental mapping of O-K, Fe-K and Ni-K edges.

As shown in the linear sweep voltammograms in Figure S5, all of the studied photoanodes exhibited a cathodic shift in the onset potential ($U_{on}$) of 110-220 mV after the deposition of FeNiO$_x$ overlayer. This result shows that the modified photoelectrodeposition process can be easily adjusted and applied to various kinds of photoanodes, including resonant light trapping architectures,[4,25] independent of initial photoelectrochemical performance and/or surface morphology. However, due to the various morphologies, substrates, and deposition processes, a direct comparison between the performances of the differently deposited photoanodes is not possible. Therefore, ultraflat (001) oriented heteroepitaxial hematite films deposited on transparent NTO coated sapphire substrates[28] were selected as a model system in order to study the effect of the FeNiO$_x$ overlayer without contributions from rough surfaces or spurious optical effects as in the case of the other samples.

Two hematite/NTO photoanodes were deposited under identical conditions, but with different dopants: 1% Sn and 1% Zn. Photoelectrochemical stability and reproducibility measurements, UV-Vis optical spectrophotometry, and XPS analysis of the heteroepitaxial hematite photoanodes on NTO before and after FeNiO$_x$ overlayer deposition are presented in the SI. It was shown that the photoelectrodeposition process is highly reproducible (Figure S6a) and the FeNiO$_x$ coated hematite proved stable during a 50 hour stability test under illumination and applied potential of 1.4 V$_{RHE}$ (Figure S6b). The transparency of the FeNiO$_x$ overlayer is demonstrated by the UV-Vis spectrophotometry measurements presented in Figure S8. Both the front and back illumination absorption spectra showed almost no change after the FeNiO$_x$ overlayer deposition. The XPS analysis presented in Figure S10 shows the Ni 2p3 peak, confirming the presence of the FeNiO$_x$ overlayer. In order to perform quantitative chemical analysis, a thick FeNiO$_x$ overlayer (~25 nm) was deposited on a heteroepitaxial Sn-doped hematite photoanode. TEM micrographs and XPS data for the thick overlayer are presented in Figures S9 and S10 respectively. The Fe:Ni cation ratio was obtained using XPS and TEM EDS methods, yielding a Fe:Ni cation ratio of 6.9:1 and 7.7:1, respectively.

The linear sweep voltammograms for both photoanodes under solar simulated illumination conditions (AM1.5G), before and after deposition of the ultrathin (~2 nm) FeNiO$_x$ overlayer, are presented in Figure 2. As can be seen for both photoanodes, there is improvement in performance after the deposition of the FeNiO$_x$ overlayer and the onset potential is cathodically shifted by approximately 100 mV. In case of the Sn-doped photoanode, the improvement is observed for the potential range of 1.05-1.50 V$_{RHE}$ whereas for higher potentials, the photocurrent remains similar to that without the FeNiO$_x$ co-catalyst. In case of the Zn-doped hematite photoanode, the improvement is observed for the whole range of potentials higher than the onset potential of 0.82 V$_{RHE}$. To investigate differences in behavior of FeNiO$_x$ overlayer on the two photoanodes, IMPS measurements were performed.[20,26,29–32]

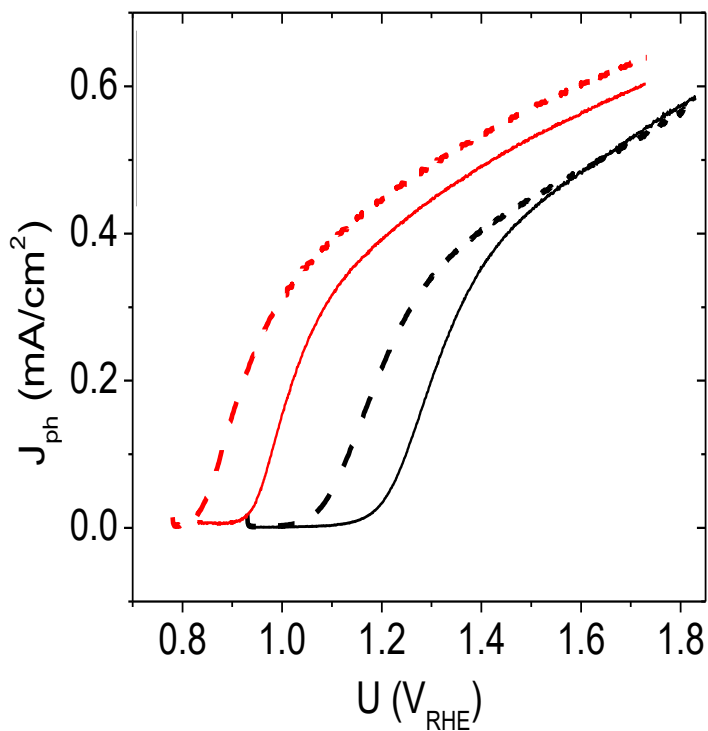

**Figure 2:** Photocurrent voltammograms obtained by linear sweep voltammetry measurements under solar-simulated light for 1% Sn- (black curves) and 1% Zn-doped (red curves) heteroepitaxial hematite photoanodes, before and after the ultrathin (~2 nm) FeNiO$_x$ overlayer deposition (solid and dashed line curves, respectively). The photocurrent density is calculated by subtracting the dark current from the light current and dividing the result by the area exposed to the light. Dark and light currents voltammograms are shown in Figure S7.

Nyquist plots of IMPS data under applied potentials ranging from 1.075 to 1.400 V$_{RHE}$ for the Sn-doped hematite photoanode before and after FeNiO$_x$ overlayer deposition are presented in Figure 3. Nyquist plots for the Zn-doped photoanode are presented in Figure 4. As can be seen, these IMPS spectra are composed of two semicircles, with high- and low-frequency intersects (HFI and LFI, respectively) with the real axis (see Figure 3(c)). It is usually assumed that the HFI or $Y_{pc}^+(0)$ is related to the flux of holes to the surface (hole current).[32,21] The LFI is supposed to account for the transferred charges at the photoelectrode/electrolyte interface, that is the hole current minus the surface recombination current of electrons from the conduction band that recombine with holes at the surface, or $Y_{pc}^-(0)$.[32] Qualitative examination of the IMPS spectra shows that $Y_{pc}^+(0)$ increases with potential whereas $Y_{pc}^-(0)$ decreases. These trends are attributed to increase of the hole current and decrease of the recombination current with an increase in applied potential.

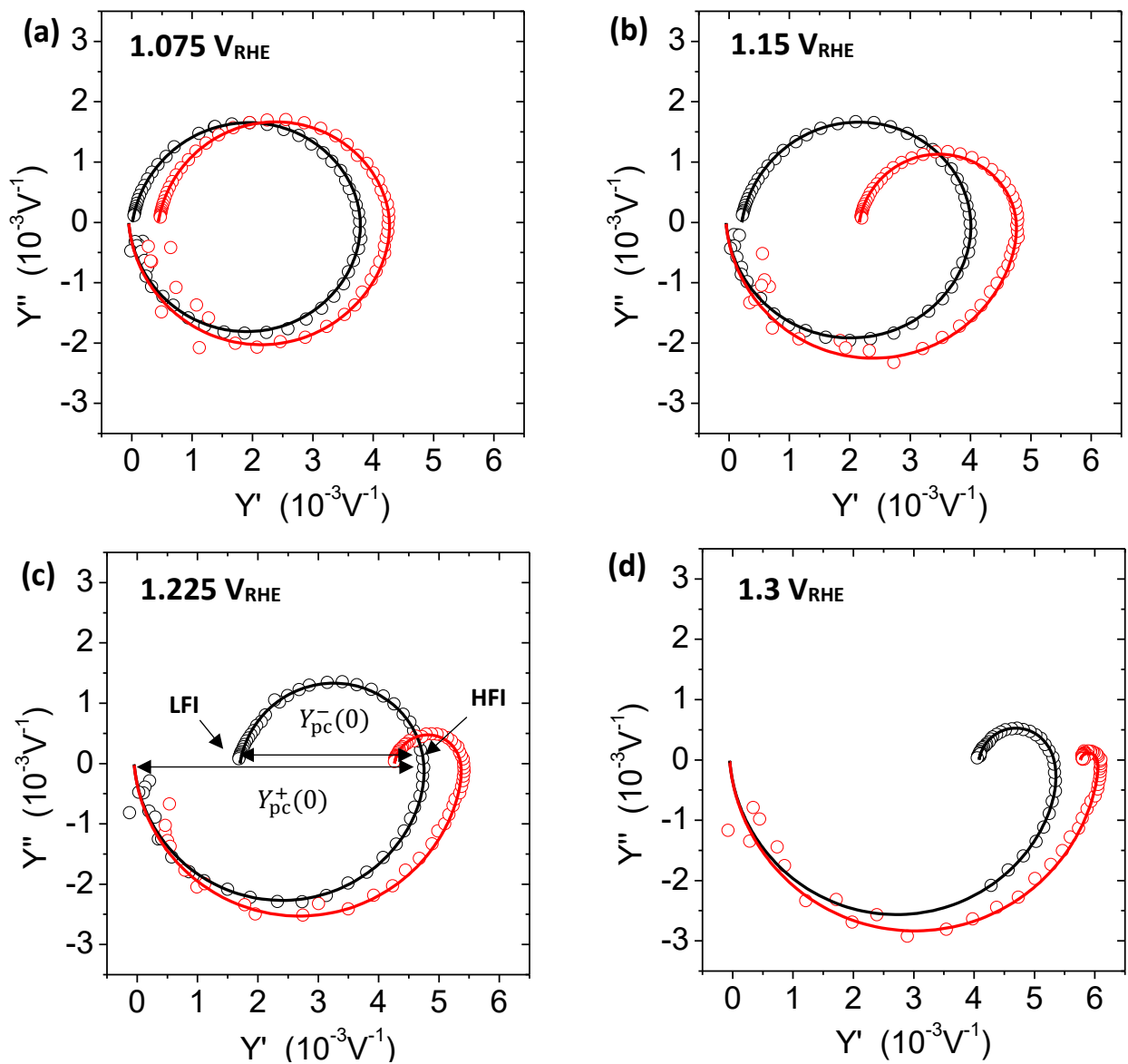

**Figure 3: IMPS spectra for 1% Sn-doped hematite sample before (black circles) and after (red circles) FeNiO$_x$ overlayer deposition, measured at potentials of (a) 1.075, (b) 1.15, (c) 1.225 and (d) 1.3 V$_{RHE}$. Solid lines represent data fitting.**

The differences in the IMPS spectra before and after the FeNiO$_x$ overlayer deposition is significant. As can be seen in Figures 3 and 4, a decrease in $Y_{pc}^-(0)$ (or the shift of LFI to higher values) is observed at all the applied potentials for both of the photoanodes. The decrease of $Y_{pc}^-(0)$ after co-catalyst deposition is more pronounced for the Zn-doped photoanode and becomes negligibly small at potentials of 1 V$_{RHE}$ and higher. For both photoanodes $Y_{pc}^+(0)$ increases after co-catalyst deposition, however this effect is more pronounced for the Zn-doped photoanode.

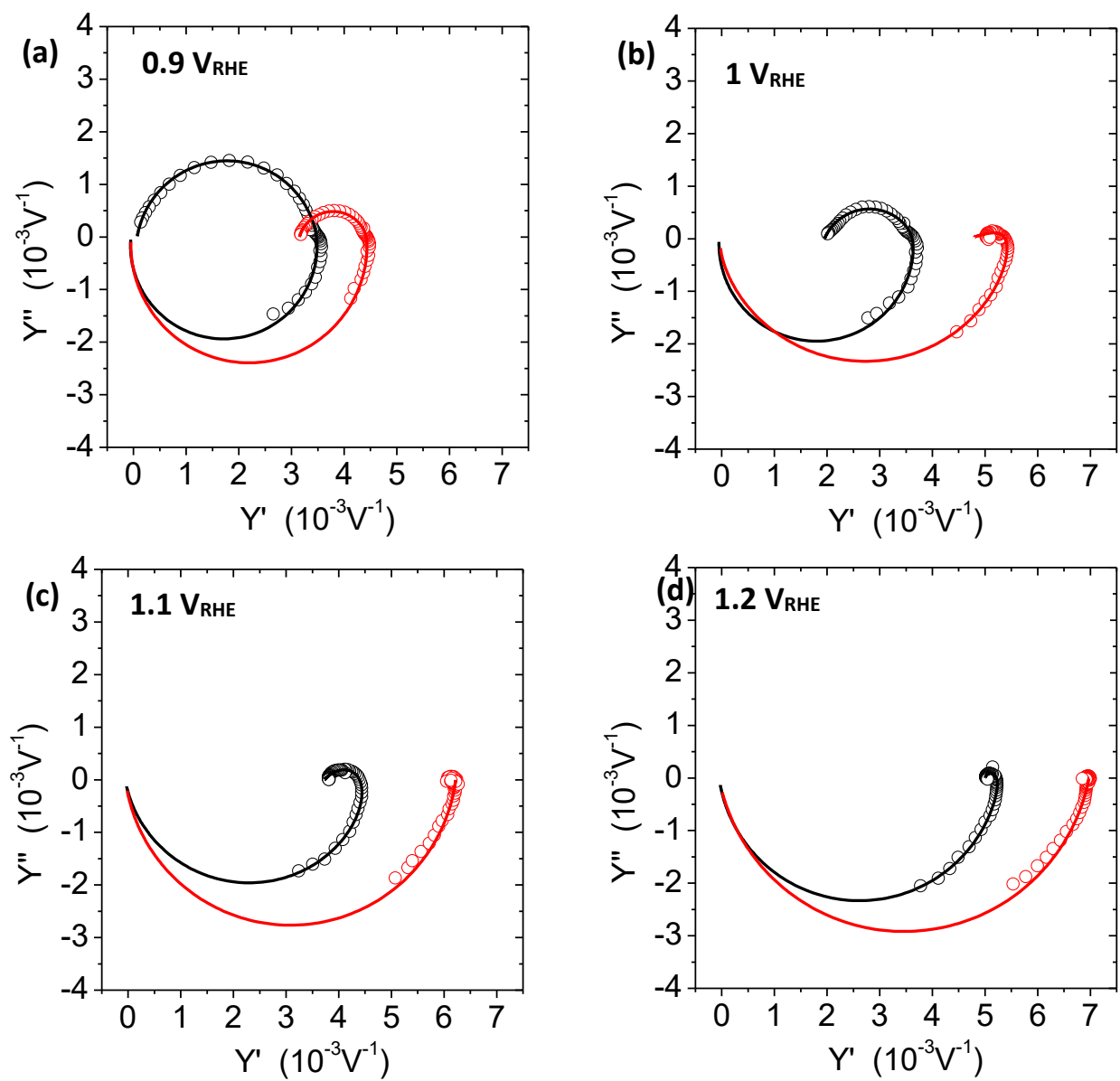

**Figure 4:** IMPS spectra for 1% Zn-doped hematite sample before (black circles) and after (red circles) FeNiO$_x$ overlayer deposition, measured at potentials of (a) 1.075, (b) 1.15, (c) 1.225 and (d) 1.3 V$_{RHE}$. Solid lines represent data fitting.

Quantitative analysis of the IMPS spectra requires careful consideration of partially overlapping semicircles.[26] Furthermore, the photocurrent and recombination current of hematite photoanodes often display a nonlinear dependence on the light intensity.[26,32] Therefore, IMPS spectra at different potentials and two different light intensities were fitted to a simple equivalent circuit model[29] (see Figure S11) in order to quantify $Y_{pc}^{+}(0)$ and $Y_{pc}^{-}(0)$. The rigorous IMPS analysis suggested in Ref. [26] was conducted, yielding the hole current ($J_h$), recombination current ($J_r$), and photocurrent ($J_{ph}$) as a function of the applied potential for a given light intensity. For a more detailed introduction into this method see Refs. [26,29].

$J_h$, $J_r$ and $J_{ph}$ for the Sn-doped and the Zn-doped hematite photoanodes without and with FeNiO$_x$ overlayers, for white LED illumination of 100 mW/cm² are presented in Figure 5a and 5b, respectively, together with the respective photocurrent voltammograms. The black circles and triangles represent $J_{ph}$ as obtained by IMPS analysis. They agree well with the respective photocurrent voltammograms measured under quasi-static conditions. As the two quantities were measured and calculated independently, the good agreement serves as validation for the IMPS analysis.

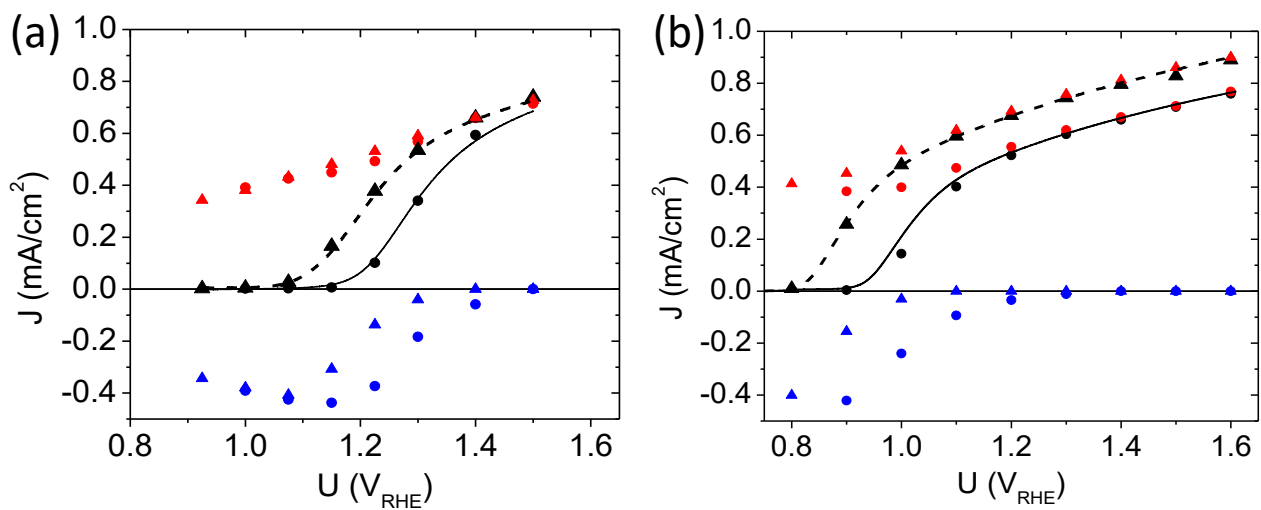

**Figure 5: Results of the IMPS analysis showing $J_h$ (hole current, red), $J_{ph}$ (photocurrent, black) and $J_r$ (surface recombination current, blue) as a function of the applied potential for 1% Sn- (a) and 1% Zn-doped (b) hematite photoanodes. Circles and triangles stand for photoanodes before and after the FeNiO$_x$ overlayer deposition, respectively. Full and dashed lines represent photocurrent voltammograms measured before and after the FeNiO$_x$ overlayer deposition, respectively. Measurements were performed in a Zahner CIMPS system with a white light LED, under illumination of 100 mW/cm².**

For the Sn-doped hematite photoanode, the IMPS analysis shows that the improvement in the photocurrent is largely due to suppression of $J_r$ and that $J_h$ remains almost unchanged after the FeNiO$_x$ overlayer deposition. This is consistent with other studies.[20,21] The reduction in surface recombination has been attributed to surface passivation,[20] hole storage,[21] improved catalysis,[17] and effective hole separation.[33] In this study we have also conducted the "traditional" IMPS analysis as proposed by Peter and co-workers.[34] The respective values of charge transfer and recombination rates are presented in Table S1, but they do not yield a definitive trend, consistent with an earlier report.[20] Additional discussion of these results is provided in the SI.

With the presented empirical IMPS analysis, we cannot distinguish between the possible mechanisms for reduced surface recombination yet, e.g. whether it is related to improved catalysis as previously reported for ultrathin photoelectrodeposited layers[17] or to other effects such as hole storage or passivation typically reported for thicker overlayers.[20,21] However, the IMPS analysis yields the magnitude of the surface recombination current without *a priori* assumption of a mechanistic model. No sacrificial reagents or complicated measurement setup are required and the related measurements can be conducted *in operando*. As can be seen in Figure 5a, the surface recombination current at the reversible potential of 1.23 $V_{RHE}$ for the Sn-doped sample is reduced from $-0.38\,\frac{mA}{cm^2}$ without the NiFeO$_x$ overlayer to $-0.10\,\frac{mA}{cm^2}$ with it. The photocurrent densities for both of the samples without and with NiFeO$_x$ co-catalyst are summarized in Table 1, which also presents the change in photocurrent for each of the measured potentials. For the Sn-doped sample, the increase in photocurrent reaches its maximum at potential ~1.225 $V_{RHE}$ and then plateaus at higher potentials, reaching a similar photocurrent value of the sample without co-catalyst. This reflects the data extracted from IMPS analysis, showing unchanged $J_h$ and suppression of $J_r$. At high potentials, the recombination is negligible and thus any change in photocurrent would originate from hole current to the surface, which was unchanged for the Sn-doped sample.

**Table 1: Photocurrents for Sn and Zn-doped samples without and with co-catalyst.**

| 1%Sn-doped hematite | | | | 1%Zn-doped hematite | | | |
|---|---|---|---|---|---|---|---|
| U ($V_{RHE}$) | $J_{ph}$ bare (mA/cm$^2$) | $J_{ph}$ with co-catalyst (mA/cm$^2$) | $\Delta J_{ph}$ (mA/cm$^2$) | U ($V_{RHE}$) | $J_{ph}$ bare (mA/cm$^2$) | $J_{ph}$ with co-catalyst (mA/cm$^2$) | $\Delta J_{ph}$ (mA/cm$^2$) |
| 1 | 0.002 | 0.004 | 0.002 | 0.9 | 0.003 | 0.257 | 0.254 |
| 1.075 | 0.003 | 0.027 | 0.024 | 1 | 0.144 | 0.486 | 0.342 |
| 1.15 | 0.006 | 0.165 | 0.159 | 1.1 | 0.402 | 0.596 | 0.194 |
| 1.225 | 0.101 | 0.377 | 0.276 | 1.2 | 0.523 | 0.675 | 0.152 |
| 1.3 | 0.340 | 0.535 | 0.195 | 1.3 | 0.603 | 0.744 | 0.141 |
| 1.4 | 0.594 | 0.660 | 0.066 | 1.4 | 0.659 | 0.795 | 0.136 |
| 1.5 | 0.727 | 0.740 | 0.013 | 1.5 | 0.708 | 0.828 | 0.12 |

Figure 5b shows the IMPS analysis of the 1% Zn-doped epitaxial film. A reduction of surface recombination is observed at low potentials, where this process is dominant. Interestingly, the analysis also reveals similar increase in $J_h$ of approximately $0.14\,\frac{mA}{cm^2}$ for all potentials higher than 1.15 $V_{RHE}$ after the FeNiO$_x$ overlayer deposition, which originates mostly from hole current as the surface recombination current is negligible.

It is worth noting that, when compared to Sn-doped photoanode, the surface recombination current of the Zn-doped photoanode vanishes at much lower potentials. This appears to be consistent with the high and low electron concentrations in the Sn- and Zn-doped hematite photoanodes,[27] respectively, that give rise to stronger recombination in the Sn-doped photoanode than in the Zn-doped photoanode. The increase in photocurrent of the Zn-doped sample at higher potentials, as presented in Table 1, originates from an increase in hole current, as demonstrated by the IMPS analysis. This is supported by complementary hole scavenger measurements (see Figure S14), which also show that the hole current does not increase after the NiFeO$_x$ overlayer deposition for the Sn-doped hematite photoanode but it does increase for the Zn-doped photoanode. The improvement in $J_h$ suggests that the FeNiO$_x$ overlayer does not only affect the properties of the surface, but also assists in charge separation in the bulk. A possible reason is related to enhanced asymmetry in the energetic profile created by the deposition of the overlayer, which creates a driving force for greater charge separation. This change can be related to unpinning of the Fermi level at the surface after the deposition of the co-catalyst overlayer, as was suggested elsewhere.[35] The effect of fermi level unpinning is expected to be more pronounced in case of the Zn-doped hematite that behaves as a weak n-type photoanode, probably due to electron injection from the NTO transparent electrode.[27] In this case, the Zn-doped hematite photoanode is most likely fully depleted with no built-in electrical field assists charge separation in donor-doped hematite photoanodes.[36] Consequently, unpinning of the Fermi level at the surface, which has been attributed to passivation of surface states by the ultrathin co-catalyst overlayer, would provide additional band bending at the surface under applied potential. In the case of the Sn-doped hematite photoanode, this effect is expected to be less pronounced as the built-in electrical field inside the depletion layer assists in charge separation in the bare hematite photoanode, without the co-catalyst. Another possible reason for the improved hole transport to the surface is related to enhanced charge transfer kinetics which would draw more holes to the surface. However, this was reported not to be the case for conventional hematite photoanodes.[37] Another possible mechanism for the performance enhancement has been previously related to the ability of thicker overlayers (~100 nm) to collect and store photogenerated holes.[21] However, this effect is unlikely in the studied case given the small thickness of the deposited co-catalyst overlayer.

As both of the examined samples were prepared under identical deposition conditions with the dopant being the only difference between them, it can be concluded that the effect of the $FeNiO_x$ overlayer does not only depend on its own composition,[21] but also on the doping of the underlying hematite layer and on interface energetics between the hematite and the $FeNiO_x$ overlayer. As a result, different conclusions regarding the role of $FeNiO_x$ layers from various studies could potentially be ascribed to differences in the hematite doping. Undoped hematite is naturally n-type, and the doping concentration is significantly affected by preparation conditions and impurities. Precise control of doping and morphology using well-controlled heteroepitaxial growth allows for systematic investigation of the role of the $FeNiO_x$ overlayers on hematite. Further investigations can optimize the hematite doping with respect to the co-catalyst, providing another route for adjustment and improvement of photoanode performance.

In conclusion, photoelectrochemical deposition of ultrathin (~2 nm) transparent $FeNiO_x$ overlayers improves the performance of both Sn- and Zn-doped hematite photoanodes. Empirical IMPS analysis of the photoanodes shows that the $FeNiO_x$ overlayer enhances the hole current to the surface and reduces the surface recombination current for the Zn-doped hematite photoanode. In contrast, for the Sn-doped photoanode, only the surface recombination current was reduced. These results show that the $FeNiO_x$ overlayer can play different roles depending on the type of doping in the photoanode it is deposited onto, suggesting that optimizing the interfacial properties between the semiconductor photoanode and co-catalyst overlayer are critical to achieve maximal improvement in performance.

**Acknowledgement**

The research leading to these results has received funding from the European Research Council under the European Union's Seventh Framework programme (FP/200702013) / ERC (grant agreement n. 617516). D.A.G. acknowledges support by Marie-Sklodowska-Curie Individual Fellowship No. 659491. K. D. Malviya acknowledges support in part at the Technion by a fellowship from the Lady Davis Foundation.

**Supporting information**

Detailed experimental procedures; hematite and $FeNiO_x$ preparation; schematics of all prepared samples; AFM and XPS characterization; optical spectra measured by UV-Vis spectrophotometer; additional photoelectrochemical measurements; hole scavenger measurements; residuals for IMPS analysis.

**References**

(1) Sivula, K.; Le Formal, F.; Grätzel, M. *ChemSusChem* **2011**, *4* (4), 432–449.

(2) Guo, X.; Wang, L.; Tan, Y. *Nano Energy* **2015**, *16*, 320–328.

(3) Peerakiatkhajohn, P.; Yun, J.-H.; Chen, H.; Lyu, M.; Butburee, T.; Wang, L. *Adv. Mater.* **2016**, *28* (30), 6405–6410.

(4) Dotan, H.; Kfir, O.; Sharlin, E.; Blank, O.; Gross, M.; Dumchin, I.; Ankonina, G.; Rothschild, A. *Nat. Mater.* **2012**, *12* (2), 158–164.

(5) Warren, S. C.; Voïtchovsky, K.; Dotan, H.; Leroy, C. M.; Cornuz, M.; Stellacci, F.; Hébert, C.; Rothschild, A.; Grätzel, M. *Nat. Mater.* **2013**, *12* (9), 842–849.

(6) Kay, A.; Cesar, I.; Grätzel, M. *J. Am. Chem. Soc.* **2006**, *128* (49), 15714–15721.

(7) Zhong, D. K.; Gamelin, D. R. *J. Am. Chem. Soc.* **2010**, *132* (12), 4202–4207.

(8) Young, K. M. H.; Hamann, T. W. *Chem. Commun.* **2014**, *50* (63), 8727–8730.

(9) Badia-Bou, L.; Mas-Marza, E.; Rodenas, P.; Barea, E. M.; Fabregat-Santiago, F.; Gimenez, S.; Peris, E.; Bisquert, J. *J. Phys. Chem. C* **2013**, *117* (8), 3826–3833.

(10) Tilley, S. D.; Cornuz, M.; Sivula, K.; Grätzel, M. *Angew. Chemie* **2010**, *122* (36), 6549–6552.

(11) Shi, Q.; Lu, R.; Lu, L.; Fu, X.; Zhao, D. *Adv. Synth. Catal.* **2007**, *349* (11–12), 1877–1881.

(12) Miller, E. L.; Rocheleau, R. E. *J. Electrochem. Soc.* **1997**, *144* (9), 3072.

(13) Smith, R. D. L.; Prévot, M. S.; Fagan, R. D.; Zhang, Z.; Sedach, P. A.; Siu, M. K. J.; Trudel, S.; Berlinguette, C. P. *Science (80-. ).* **2013**, *340* (6128).

(14) Lu, X.; Zhao, C. *Nat. Commun.* **2015**, *6*, 6616.

(15) Liu, K.-C.; Anderson, M. A. *J. Electrochem. Soc.* **1996**, *143* (1), 124.

(16) Beverskog, B.; Puigdomenech, I. *Corros. Sci.* **1997**, *39* (5), 969–980.

(17) Morales-Guio, C. G.; Mayer, M. T.; Yella, A.; Tilley, S. D.; Grätzel, M.; Hu, X. *J. Am. Chem. Soc.* **2015**, *137* (31), 9927–9936.

(18) Du, C.; Yang, X.; Mayer, M. T.; Hoyt, H.; Xie, J.; McMahon, G.; Bischoping, G.; Wang, D. *Angew. Chemie Int. Ed.* **2013**, *52* (48), 12692–12695.

(19) Li, J.; Meng, F.; Suri, S.; Ding, W.; Huang, F.; Wu, N. *Chem. Commun. Chem. Commun* **2012**, *48* (48), 8213–8215.

(20) Thorne, J. E.; Jang, J.-W.; Liu, E. Y.; Wang, D. *Chem. Sci.* **2016**, *7* (5), 3347–3354.

(21) Hajibabaei, H.; Schon, A. R.; Hamann, T. W. *Chem. Mater.* **2017**, *29* (16), 6674–6683.

(22) Zandi, O.; Hamann, T. W. *Phys. Chem. Chem. Phys.* **2015**, *17* (35), 22485–22503.

(23) Tamirat, A. G.; Su, W.-N.; Dubale, A. A.; Chen, H.-M.; Hwang, B.-J. *J. Mater. Chem. A* **2015**, *3* (11), 5949–5961.

(24) Qiu, J.; Hajibabaei, H.; Nellist, M. R.; Laskowski, F. A. L.; Hamann, T. W.; Boettcher, S. W. *ACS Cent. Sci.* **2017**, *3* (9), 1015–1025.

(25) Grave, D. A.; Dotan, H.; Levy, Y.; Piekner, Y.; Scherrer, B.; Malviya, K. D.; Rothschild, A. *J. Mater. Chem. A* **2016**, *4* (8), 3052–3060.

(26) Klotz, D.; Ellis, D. S.; Dotan, H.; Rothschild, A. *Phys. Chem. Chem. Phys.* **2016**, *18* (34), 23438–23457.

(27) Kay, A.; Grave, D. A.; Ellis, D. S.; Dotan, H.; Rothschild, A. *ACS Energy Lett.* **2016**, *1* (4), 827–833.

(28) Grave, D. A.; Klotz, D.; Kay, A.; Dotan, H.; Gupta, B.; Visoly-Fisher, I.; Rothschild, A. *J. Phys. Chem. C* **2016**, *120* (51), 28961–28970.

(29) Klotz, D.; Grave, D. A.; Rothschild, A. *Phys. Chem. Chem. Phys.* **2017**, *19* (31), 20383–20392.

(30) L. Dloczik, †; O. Ileperuma, ‡; I. Lauermann, †; L. M. Peter, *; E. A. Ponomarev, ‖; G. Redmond, §; N. J. Shaw, § and; Uhlendorf†, I. *J. Phys. Chem. B* **1997**, *101* (49), 10281–10289.

(31) Ponomarev, E. A.; Peter, L. M. *J. Electroanal. Chem.* **1995**, *396* (1–2), 219–226.

(32) Peter, L. M.; Wijayantha, K. G. U.; Tahir, A. A.; Wijayantha, K. G. U.; Gratzel, M.; Sivula, K.; Frydrych, J.; Gratzel, M. *Faraday Discuss.* **2012**, *155* (0), 309–322.

(33) Mcdonald, K. J.; Choi, K.-S. *Chem. Mater* **2011**, *23*, 1686–1693.

(34) Peter, L. M. *J. Solid State Electrochem.* **2013**, *17* (2), 315–326.

(35) Thorne, J. E.; Li, S.; Du, C.; Qin, G.; Wang, D. *J. Phys. Chem. Lett.* **2015**, *6* (20), 4083–4088.

(36) Orton, J. W.; Powell, M. J. *Reports Prog. Phys.* **1980**, *43* (11), 1263–1307.

(37) Barroso, M.; Mesa, C. A.; Pendlebury, S. R.; Cowan, A. J.; Hisatomi, T.; Sivula, K.; Grätzel, M.; Klug, D. R.; Durrant, J. R. *Proc. Natl. Acad. Sci. U. S. A.* **2012**, *109* (39), 15640–15645.